\documentclass[prb,twocolumn,superscriptaddress,noshowpacs]{revtex4}
\usepackage{graphicx}
\usepackage{physics}
\usepackage{bm}
\usepackage{ulem}
\usepackage{color}
\newcommand{\tcr}[1]{\textcolor{black}{#1}}

\begin{document}

\title{Kibble-Zurek mechanism in polariton graphene}

\author{D.~D.~Solnyshkov}
\affiliation{Institut Pascal, PHOTON-N2, Universit\'e Clermont Auvergne, CNRS, SIGMA Clermont, F-63000 Clermont-Ferrand, France.}
\affiliation{Institut Universitaire de France (IUF), F-75231 Paris, France}

\author{L.~Bessonart}
\affiliation{Institut Pascal, PHOTON-N2, Universit\'e Clermont Auvergne, CNRS, SIGMA Clermont, F-63000 Clermont-Ferrand, France.}

\author{A. Nalitov}
\affiliation{Institut Pascal, PHOTON-N2, Universit\'e Clermont Auvergne, CNRS, SIGMA Clermont, F-63000 Clermont-Ferrand, France.}
\affiliation{Faculty of Science and Engineering, University of Wolverhampton, Wulfruna St, Wolverhampton WV1 1LY, UK}

\author{G.~Malpuech}
\affiliation{Institut Pascal, PHOTON-N2, Universit\'e Clermont Auvergne, CNRS, SIGMA Clermont, F-63000 Clermont-Ferrand, France.}

\begin{abstract}
We study the formation of topological defects (quantum vortices) \tcr{during the formation} of a 2D polariton condensate at the $\Gamma$ point of a honeycomb lattice via the Kibble-Zurek mechanism. \tcr{The lattice modifies the single-particle dispersion. The typical interaction energies at the quench time correspond to the linear part of the dispersion. The resulting scaling exponent for the density of topological defects is numerically found as $0.95\pm0.05$. This value differs from the one expected for 2D massive particles (1/2), but is indeed compatible with the one expected for a linear dispersion}. We moreover demonstrate that the vortices can be pinned to the lattice, which prevents their recombination and could facilitate their observation and counting in \tcr{continuous wave} experiments. 
\end{abstract}


\maketitle 

\section{Introduction}
Topological photonics is a rapidly developing area which already covers a wide range of hot topics of modern physics \cite{lu2014topological,Ozawa2019}. Interesting non-trivial effects with important practical applications, such as the quantum Hall effect (normal and anomalous) and the associated chiral edge states that can be used for topological lasing \cite{Solnyshkov2016,St-Jean2017,Bahari2017,Bandres2018,zeng2020electrically} or optical isolation \cite{Solnyshkov2018}, were observed in photonic lattices. Photonic systems have allowed measuring the Berry curvature \cite{gianfrate2020measurement} characterizing such topological effects.

Among different photonic platforms, exciton-polaritons (polaritons) \cite{Microcavities} claim a special place thanks to their strong intrinsic non-linearity. Indeed, they arise from strong light-matter coupling, and as such can benefit from the repulsive exciton-exciton interactions, ensuring an effective $\chi^3$ coefficient $10^6$ times higher than the typical Kerr nonlinearity \cite{emmanuele2020highly}. The  polariton platform is also extremely practical from the experimental point of view, offering full access to the system's wavefunction (density and phase) control and measurement both in the real space and in the reciprocal space.

These unique features of exciton-polaritons allow merging topological photonics \cite{lu2014topological} with the field of interacting quantum fluids \cite{Smirnova2020,solnyshkov2020microcavity}.
The first important question arising at the crossroads of the two fields is the modification of topology by interactions of the fluids \cite{Furukawa2015,Bleu2017x,Sigurdsson2019} and, in particular, the stability of the topologically protected modes in lasers \cite{Kartashov2019,Secli2019,Zapletal2020}.
The possibilities of synergy of the two fields were already demonstrated by enhancement of topological protection in lattices due to quantum fluid interactions \cite{Bleu2018nc}.

Aside from the band topology context, quantum fluids  have been known since the dawn of their study to support topological defects \cite{LeggettBook}, the most famous example being the quantum vortex \cite{Pitaevskii}. Quantum vortices are protected by a particular real-space topological invariant known as the winding number \cite{Thouless1998}, which ensures their stability. They can only be removed via vortex-antivortex annihilation, or by reaching the zero-density system boundaries. Quantum vortices have been observed in liquid Helium \cite{Onsager1949,Feynman1956}, in atomic condensates \cite{Ketterle2001}, in light beams propagating in atomic vapors \cite{Zhang2019}, and also in polaritonic systems \cite{lagoudakis2008quantized}. Here, quantum vortices can be created in non-equilibrium flows via quantum turbulence mechanisms \cite{sanvitto2010persistent,nardin2011hydrodynamic}. Another way is based on preferential condensation at selected orbital angular momentum states \cite{Sala2015,zambon2019optically,Yulin2020}. It is also possible to imprint quantum vortices at will under resonant pumping \cite{boulier2015vortex,dominici2018interactions}, using spatial light modulators.

A particularly interesting mechanism of generation of topological defects is the Kibble-Zurek mechanism (KZM), based on the intrinsic non-adiabaticity of second-order phase transitions. Initially proposed in the context of the spontaneous symmetry breaking defects in the early Universe evolution by Kibble \cite{Kibble1976}, it was later applied to determine the residual quantum vortex density in superfluid liquid Helium by Zurek \cite{Zurek1985,Zurek1996}, and then generalized to other systems and configurations \cite{campo2014universality,yukalov2015}. KZM is characterised by a particular scaling of the density of the topological defects as a function of the quenching speed (the characteristic time of the phase transition). For homogeneous polariton condensates (without lattices), these scalings were studied in several recent theoretical works \cite{Mat2014,Matu2017,Comaron2018}. One of the most important problems for the experimental studies of this mechanism is the recombination of the topological defects (vortex-antivortex recombination): if they are allowed to move freely and do not benefit from any additional protection, their density decreases with time after the transition has occurred (as already noticed by Kibble\cite{Kibble1976}), and drops to zero at long time scales. Recently, it was shown how this problem can be solved by additional topological protection for the defects in 1D polariton lattices \cite{Solnyshkov2016}. KZM in lattices was the subject of several recent studies \cite{dora2019kibble,keesling2019quantum}, some of which were focused on Dirac points \cite{dutta2010quenching,Lledo2021} and topological insulators \cite{Lara2020}. \tcr{A specificity brought by lattices with respect to homogeneous system is the change of the dispersion, whose non-parabolicity can affect the scaling exponents. The honeycomb lattice is particularly interesting among others, because it can exhibit a non-trivial topology if the gap is opened at the Dirac point, which can be done, for example, by using the spin-anisotropic interactions \cite{Bleu2017x}.}

In this work, we address polariton condensation at \tcr{the lowest energy state (the $\Gamma$ point)} of a honeycomb lattice of micropillar cavities.  
We demonstrate that the vortices formed via the KZM can be pinned to the potential of the lattice. This prevents their recombination and allows their observation and counting in cw \tcr{(continuous wave)} single-switch-on experiments. \tcr{We demonstrate that the density of quantum vortices exhibits a power-law scaling with the pumping power, with the exponent corresponding to the mean-field predictions for a 2D system with a linear dispersion, which in our case is induced by the lattice.}

\section{Model}

We simulate the polariton condensation under non-resonant pumping and the polariton dynamics in the honeycomb lattice using the Gross-Pitaevskii equation with lifetime, energy relaxation, and saturated gain
\begin{eqnarray}
    i\hbar\frac{\partial\psi}{\partial t}&=&-\left(1-i\Lambda\right)\frac{\hbar^2}{2m}\Delta\psi
    +g\left|\psi\right|^2\psi \nonumber\\
    &+&\left(U+i\gamma(n_{tot})-\frac{i\hbar}{2\tau}\right)\psi+\chi
    \label{gpe}
\end{eqnarray}
Here, $m$ is the polariton mass, $g$ is the polariton-polariton interaction constant, $U$ is the potential forming the honeycomb lattice of polariton graphene, $\gamma(n)$ is the saturated gain term, $n_{tot}$ the total polariton density, $\tau$ is the polariton lifetime, $\chi$ is the noise describing the spontaneuous scattering from the excitonic reservoir, and $\Lambda$ characterizes the efficiency of the energy relaxation \cite{Pitaevskii58}. We solve Eq.~\ref{gpe} numerically using the 3rd-order Adams method for the time derivative and a GPU-accelerated FFT for the Laplacian. We have chosen the parameters of a typical polariton graphene lattice \cite{Jacqmin2014} that we have already used to describe the polariton condensation in such lattice at the top of the first band. However, here we have increased the energy relaxation constant $\Lambda$ and replaced the localized Gaussian pumping of Ref.~\cite{Jacqmin2014} by a homogeneous pumping. 
Indeed, the numerous studies on the dynamics of polariton condensation have demonstrated that depending on the parameters of the system, in particular, on the exciton-photon detuning and on the size of the pumping spot, the condensation can occur either as an equilibrium thermodynamic process or as an out-of-equilibrium dynamical process \cite{Kasprzak2008,Levrat2010,Feng2013}. Increasing the detuning and the size of the pumping spot makes polariton energy relaxation more efficient. The transition between the condensation in excited states and in the ground state with these parameters has been demonstrated experimentally in a homogeneous 2D system \cite{Wertz2009}.
 
\begin{figure}[tbp]
    \centering
    \includegraphics[width=0.99\linewidth]{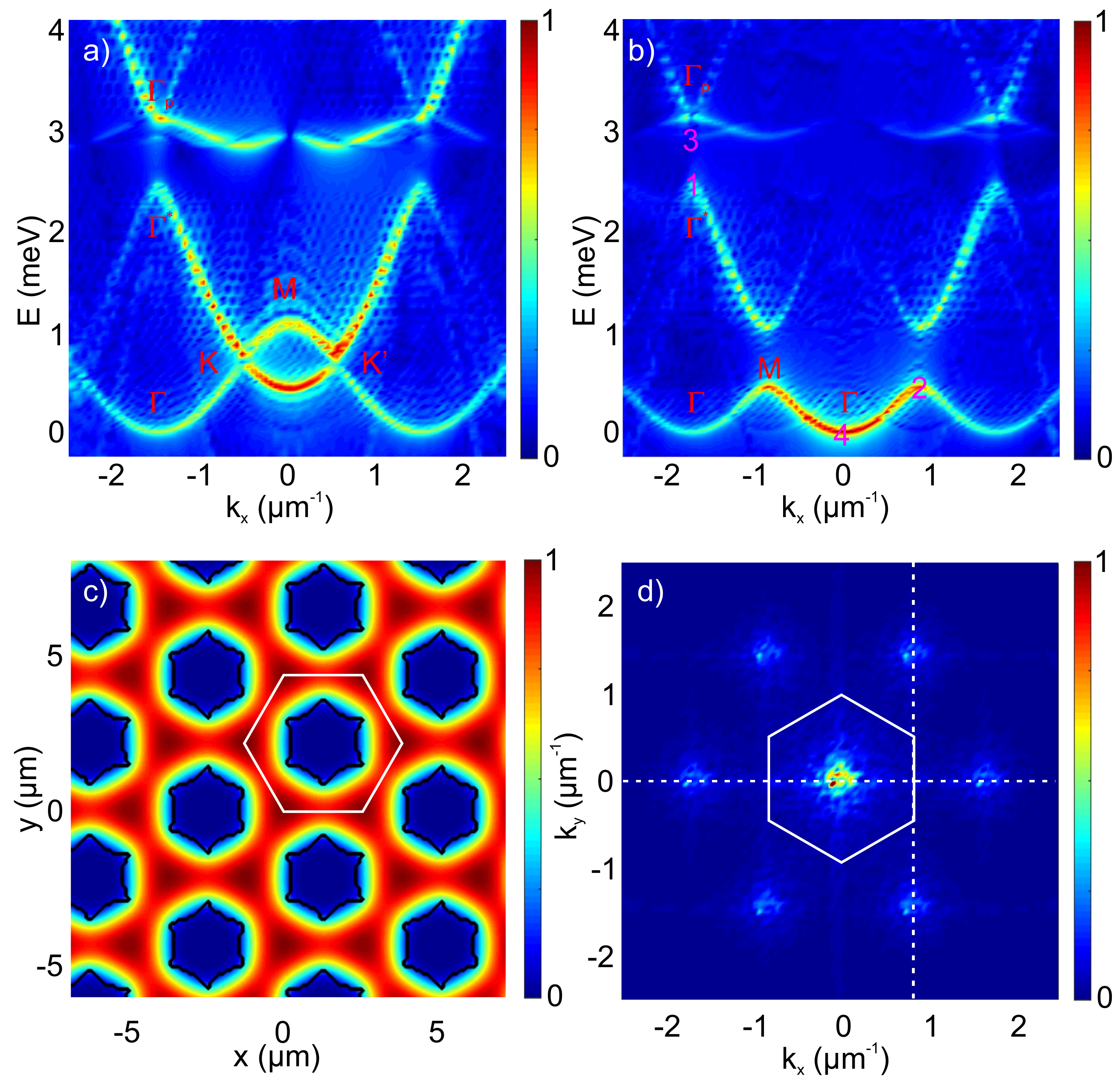}
    \caption{a,b) The numerically calculated energy dispersions of the polariton graphene in the directions (a) $k_y$ ($\Gamma\to K\to M\to K'$) and $k_x$ ($\Gamma\to M\to\Gamma$). Magenta numbers mark the condensation points in Refs.~\cite{Jacqmin2014,klembt2018exciton,Suchomel2018} and in the present work, respectively. c) Spatial image of the lattice potential and the density of the condensate formed at the $\Gamma$ point. A unit cell of the lattice is shown with white lines.  d) Reciprocal space image confirming the condensation at the $\Gamma$ point of the first Brillouin zone. White dashed lines mark the cuts shown in panels (a,b).}
    \label{fig1}
\end{figure}

 Figure~\ref{fig1}(a) shows the numerically calculated dispersion of the polariton graphene (without the condensate) in the direction $k_y$ ($\Gamma\to K \to M \to K' \to \Gamma$, marked with red characters), with well-resolved energy bands (the whole $s$ band and a part of the $p$ band). This dispersion quite accurately  reproduces the experimental observations for polariton graphene \cite{Jacqmin2014,klembt2018exciton}.  The numerical dispersion is obtained by solving Eq.~\ref{gpe} in linear regime, keeping only the kinetic energy and the potential of the lattice. Panel (b) shows the cut in the direction $k_x$ ($\Gamma\to M$). The band extrema where the polariton condensation has been observed experimentally are marked with magenta numbers from $1$ to $3$ (corresponding to Refs.~\cite{Jacqmin2014,klembt2018exciton,Suchomel2018}). One can see that it is possible to observe condensation not only in the negative mass states (favored by localized pumping), but also in positive mass states, as in Ref.~\cite{Suchomel2018}. Condensation has also been observed in positive mass states at the Dirac point of a \emph{triangular} polariton lattice \cite{Kim2013}.
 Number $4$ marks the theoretical result of the present work. \tcr{We have decided to focus at the lowest energy state, because it provides a direct comparison to a system without a lattice. It will also serve as a reference point for the future studies of KZM in the condensation at the other particular points, especially the $K$ point.}
 
 We consider the case, where polariton condensation occurs as a dynamical process, when the effective temperature of the system (determined by the parameter $\Lambda$) is kept constant, whereas the density of the particles increases over time because of the constant pumping from the reservoir (described by the term $\gamma$), which overcomes the losses.  Figure~\ref{fig1}(c) shows the contour of the lattice potential (black line), with the hexagonal unit cell shown with white lines. The spatial distribution of the density of the condensate $n(x,y)=|\psi(x,y)|^2$ is plotted using a normalized false color scale.
 Finally, Fig.~\ref{fig1}(d) shows the reciprocal space image of the polariton condensate, confirming that if the energy relaxation is sufficiently strong, the condensate can form at the lowest energy state: the $\Gamma$ point of the first Brillouin zone. This is also confirmed by the spatial distribution of the density of the condensate: the condensate shown in Fig.~\ref{fig1}(a) exhibits relatively weak density modulations, contrary to the Bloch state of the highest energy state of this band (where the condensation has  been observed in Ref.~\cite{Jacqmin2014}), which is completely antisymmetric, changes sign between the sites of the lattice, and therefore exhibits zero density between the sites.

\section{Results and discussion}

Now that we have demonstrated the possibility of the formation of the condensate at the $\Gamma$ point of the honeycomb lattice, we begin studying the formation of the topological defects via the Kibble-Zurek mechanism and their behavior after the condensation. As explained above, according to the Ginzburg-Landau theory of the phase transitions, in any second-order phase transition the relaxation time diverges when approaching the transition point. At the same time, the characteristic size of the fluctuations diverges as well, because their energy cost tends to zero. For quantum fluids described by a wave function, such domains (formed within the normal phase) correspond to the domains of coherence of the phase. The phase of the wavefunction is constant within a single domain and different between the domains. The lines of the phase jumps terminate on topological defects which are quantum vortices, whose density is proportional to the density of the phase domains and inversely proportional to their size. The details of the calculation of the defect density will be given below, because first we would like to discuss the dynamics of the defects after the condensation and the possibilities of their experimental observation.

\begin{figure}[tbp]
    \centering
    \includegraphics[width=0.99\linewidth]{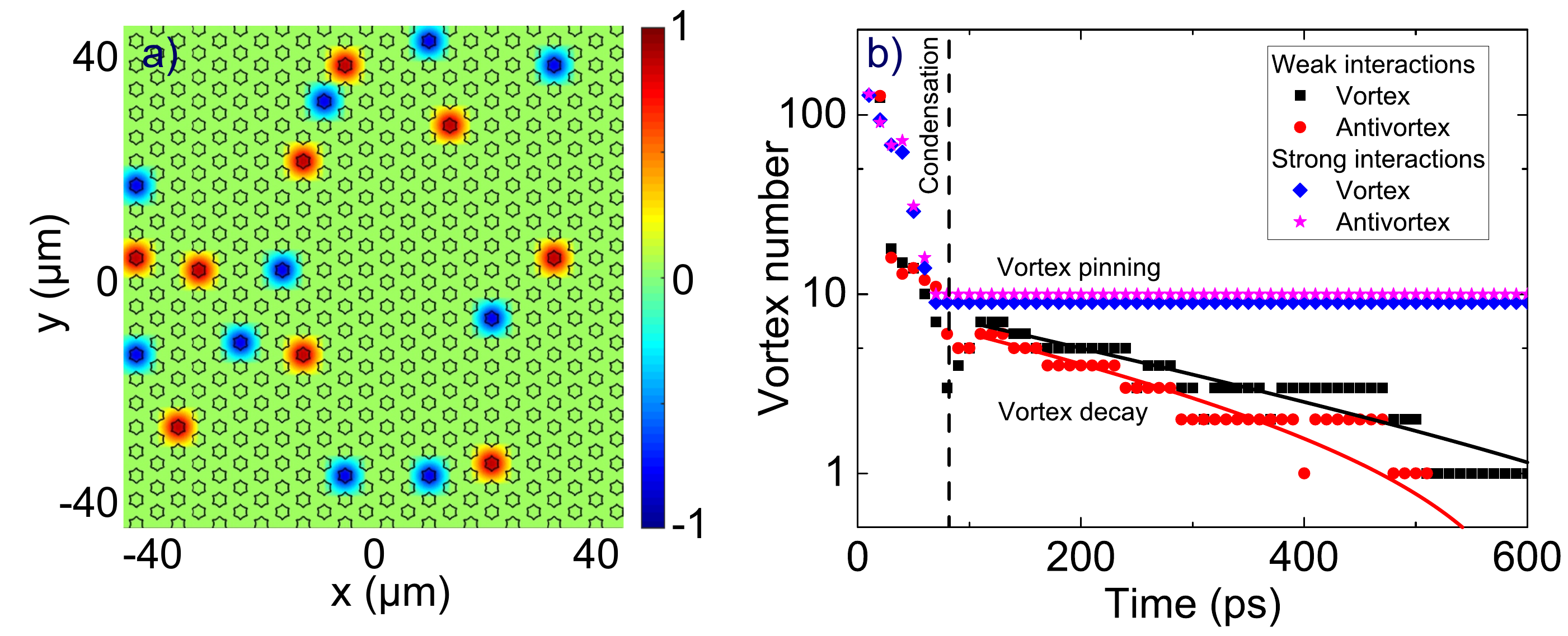}
    \caption{a) Spatial image of the curl of the wavefunction after condensation exhibiting topological defects. The lattice potential is shown as a black line. b) The number of topological defects as a function of time in the case of strong ($g n=1$~meV) and weak ($g n=0.25$~meV) interactions (points -- pinned and free vortices, respectively). Solid lines show exponential decay fits.}
    \label{fig2}
\end{figure}

It is difficult to detect vortices by studying only the density of the polariton condensate (such as shown in Fig.~\ref{fig1}(a)), especially since the vortices are attracted by low-density regions, which are the centers of the unit cells. We therefore use the curl of the condensate wave function $\mathbf{\nabla}\times \psi$ to track their positions. Indeed, since the quantum fluid described by a complex-valued wavefunction is irrotational, the only points with non-zero curl are the centers of the vortices with zero density \cite{LeggettBook}. Figure~\ref{fig2} shows the spatial image of the curl of the condensate wavefunction with a certain number of vortices (red spots) and anti-vortices (blue spots). This type of image can be observed immediately after the condensation, and the density of vortices is expected to depend on the quench time via the Kibble-Zurek mechanism. However, before studying this dependence, we will focus on the dynamics of these vortices after the condensation.

Depending on the conditions (in particular, on the characteristic interaction energy after the condensation), these vortices can either freely migrate in the system or become immediately pinned to the potential of the lattice (felt by the vortices via the density of the condensate). In the first case, the vortices and anti-vortices, exhibiting random walks, eventually approach each other or the boundary of the lattice and annihilate. Phase ordering kinetics in such conditions was studied recently for polariton condensates in Ref.~\cite{Mat2017phase}. The number of vortices (shown in Fig.~\ref{fig2}(b) with black and red points) decreases over time. \tcr{This decay is expected to be hyperbolic at high densities \cite{vinen1957mutual} (due to vortex-vortex interaction) and then turn to exponential (due to the decay on the boundaries).}
An exponential decay fit (solid black and red lines) gives a characteristic time of about $330\pm 20$~ps. It means that in a cw experiment all vortices formed during the condensation via the Kibble-Zurek mechanism will disappear before the observation (whose typical time scale is measured in seconds, and not in picoseconds). 

However, we have found that the vortices are pinned to the minima of the density (which are at the centers of the unit cells) if the characteristic interaction energy $g n$ \tcr{(with $n$ being the maximal value of the density observed at the centers of the lattice sites)} is sufficient. \tcr{Vortex pinning is a well-known phenomenon, which has enabled the first experimental observation of a quantum vortex in liquid helium \cite{vinen1958detection}. It also occurs in superconductors \cite{blatter1994vortices} and in atomic condensates \cite{Tung2006}. However, in most cases vortices are pinned to defects. Pinning to a lattice requires the vortex size $\xi=\hbar/\sqrt{2gnm}$ to be comparable with the lattice parameter $a$. For atomic condensates in optical lattices, the pinning transition was considered theoretically \cite{Pu2005,Tsubota2006} and demonstrated experimentally \cite{Tung2006}. The possibility to prevent the vortex decay in the Kibble-Zurek mechanism by the pinning was suggested by Zurek  \cite{Zurek1996} for superconductors in 2D. We note that the boundary of the system not only attracts vortices (which could pin them), but also acts as a source of decay, as a region of zero density where the phase is not defined.}
When the vortices are pinned, the vortex mutual annihilation is effectively suppressed. The condition for the transition between the two regimes is $g n\sim t$, where $t$ is the characteristic width of the energy band.  With the parameters of a typical GaAs-based polariton graphene $t\approx 0.25$~meV\cite{Jacqmin2014}, we find a critical interaction energy of the order of 0.5~meV.  \tcr{Above this value, the vortices are pinned to the lattice.} In this case, the vortex lifetime is infinite, their numbers do not change over time (blue and cyan points \tcr{in Fig.~\ref{fig2}(b)}), and they can be counted even in a cw experiment by self-interference measurements \cite{Sala2015}, because they are completely pinned and do not move at all. The possibilities to stabilize vortices in polariton condensates in inhomogeneous systems were found previously experimentally and theoretically in Refs.~\cite{sanvitto2010persistent,Ma2017vort}. 

Now that we have established the possibility of the experimental measurement of the scaling and demonstrated the formation of the topological defects in numerical experiments, let us discuss the scaling of their density that one can expect to observe.
In Kibble-Zurek mechanism, the scaling of the density of the topological defects with the dimensionless temperature $\epsilon$ in vicinity of the transition is determined by two scaling exponents: $\nu$, the critical exponent of the correlation length, and $z$, the dynamic critical exponent \cite{campo2014universality}.  The correlation length writes
\begin{equation}
    \xi\left(\epsilon\right)=\frac{\xi_0}{|\epsilon|^\nu}
\end{equation}
while the equilibrium relaxation time $\tau$ writes
\begin{equation}
    \tau\left(\epsilon\right)=\frac{\tau_0}{|\epsilon|^{z\nu}}
    \label{enrelt}
\end{equation}
The key feature of the condensation in the lattice of polariton graphene is that the critical exponent of the correlation length $\nu$ changes with respect to the condensation in a system without a lattice. 
\tcr{Indeed, it is the dispersion of an empty (bare) system (before the formation of the condensate), which determines the scaling in the Kibble-Zurek mechanism. However, it is the dispersion at the interaction energy $gn$ corresponding to the moment of the loss of adiabaticity which has to be taken into account. For a very low density and interaction energy $gn\ll t$ at the transition, one could expect to observe the same scaling as predicted for a homogeneous 2D system ($\nu=1/2$). However, in realistic experiments requiring vortex pinning, but also a reasonable number of topological defects in a finite-size sample, $gn\sim t$, as discussed above. The dispersion of polariton graphene is not parabolic, but linear in a broad range close to this value, which allows writing the following expression for the correlation length}
\begin{equation}
    \hbar c_0 \frac{1}{\xi}\sim g n \sim \epsilon \iff \xi=\frac{\xi_0}{|\epsilon|^1}
\end{equation}
giving the critical exponent $\nu=1$. We note that it is not the dispersion of the weak excitations of the condensate which needs to be used to determine the correlation length, because the defect density is determined at the moment of the loss of adiabaticity during the quench, when the extended condensate is not formed yet, and thus the low-wavevector (long wavelength) bogolons cannot exist. \tcr{We also note that the association of the vortex pinning regime with the linear dispersion at the interaction energy is not a coincidence or a peculiarity of the honeycomb lattice. It can be expected to occur in any strong lattice well-described by the tight-binding approximation, where the dispersion is always described by cosine functions $E(k)\sim \cos ka$, and the condition for the vortex localization $\xi\sim a$ (see above) qualitatively corresponds to the region with a vanishing second derivative $\partial^2 E(k)/\partial k^2\approx 0$ and a linear dispersion.}

\begin{figure}[tbp]
    \centering
    \includegraphics[width=0.99\linewidth]{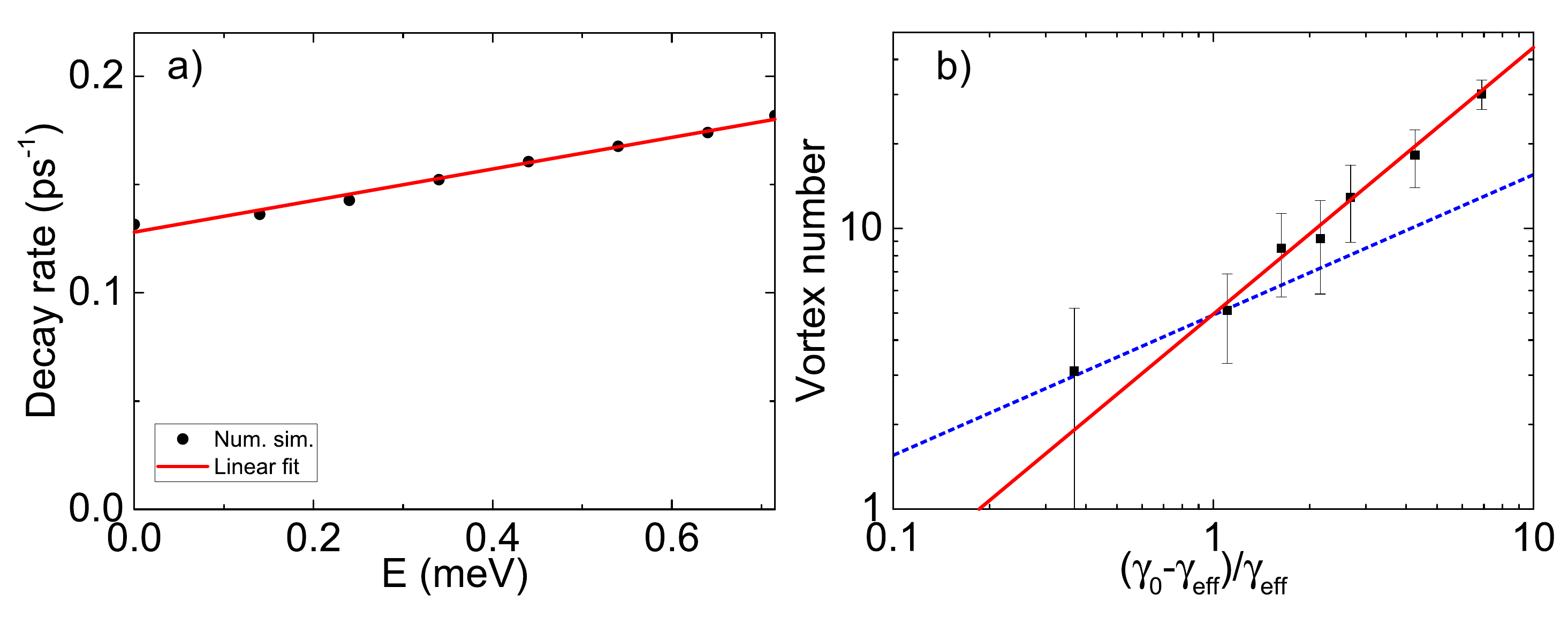}
    \caption{(a) Effective decay rate scales linearly with mode energy (circles -- numerical simulation, solid line -- linear fit). (b) Density of topological defects as a function of the quenching parameter (effective pumping) and a power law fit with a scaling exponent $0.95\pm0.05$.}
    \label{fig3}
\end{figure}

The dynamical critical exponent, linked with the energy relaxation, is controlled by the energy relaxation term $~i\Lambda\Delta\psi$ in Eq.~\eqref{gpe}. In order to determine how exactly the corresponding decay rate scales with mode energy, we have performed numerical simulations in the absence of the condensate (as for the dispersions shown in Fig.~\ref{fig1}) using relatively long pulses for the excitation of the system. The results are shown in Fig.~\ref{fig3}(a): the decay rate depends linearly on the energy of the mode in a wide range of values up to the Dirac point ($E\approx 0.7$~meV), which means that the relaxation time in Eq.~\eqref{enrelt} diverges hyperbolically, and the product $z\nu=1$, which gives $z=1$ (and not $z=2$).

The final expression for the density of topological defects contains also the dimensionality of the space $D=2$ and the dimensionality of the topological defects $d=0$. It reads
\begin{equation}
    n_{vort}=\left(\frac{\tau_0}{\tau_q}\right)^{(D-d)\frac{\nu}{1+z\nu}}=\left(\frac{\tau_0}{\tau_q}\right)^1
\end{equation}
The final scaling exponent is therefore equal to $1$. 

Changing the system's temperature with time is not the only way to vary the dimensionless temperature $\epsilon=(T-T_c)/T_c$ \tcr{and to cross the transition point $\epsilon=0$}. While it is indeed possible to observe polariton condensation by varying temperature, this variation occurs at macroscopic time scales and its speed is difficult to control. Another option is much better suited to this system: one can vary the critical temperature $T_c$ by varying the particle density $n$, because $T_c\approx \hbar^2 n/2mk_B$ (in 2D quasi-condensation). When the amplification exceeds the losses, the particle density behaves as $n(t)\sim \exp(\Gamma t)$. Linearizing this dependence at the condensation threshold density, one obtains a linear behavior for the dimensionless temperature $\epsilon(t)=t/\tau_q$, with the quench time $\tau_q\sim \Gamma^{-1}$ and $\Gamma\sim(P-P_c)$. The quench time is therefore inversely proportional to the pumping intensity excess over the threshold value. In terms of Eq.~\eqref{gpe}, the relevant parameter can be written as $(\gamma_0-\gamma_\text{eff})/\gamma_\text{eff}$, where $\gamma_0$ is the low-density limit of the saturated gain $\gamma(n)$ and $\gamma_\text{eff}$ are the total losses determined by the polariton lifetime $\tau$ and the energy relaxation $\Lambda$. Taking into account the scaling exponent found above and the fact that the vortex pinning occurs for interaction energies $g n\approx J$, we can expect that at the pumping powers corresponding to vortex pinning, the density of topological defects should increase linearly with the pumping power (scaling exponent $1$).

To test these analytical results we have performed a set of numerical experiments changing the gain $\gamma_0$, which is equivalent to changing the quench time in the Kibble-Zurek mechanism\tcr{, as discussed above}. The results are plotted in Fig.~\ref{fig3}(b), with the number of vortices observed in a system of the size $120~\mu\mathrm{m}\times 120~\mu\mathrm{m}$ shown as black points with error bars. Each point is an average over ten simulations and the error bar correspond to the root-mean-square deviation. The red line is a power law fit, which shows that the density of topological defects in our numerical simulations exhibits a scaling exponent of $0.95\pm 0.05$. This value corresponds to the analytical predictions of a mean-field theory in 2D with a linear dispersion. As explained above, the dispersion of graphene is approximately linear in the region corresponding to the critical interaction energy of the vortex pinning.The scaling behavior is different from what was observed theoretically for a 1D system with a lattice in Ref.~\cite{Solnyshkov2016}, where the protection against the decay of the topological defects was ensured not by pinning, but by a topological invariant (the Zak phase). Low density regime was accessible in that case, and the scaling exponent was corresponding to the mean-field predictions for a parabolic dispersion in 1D. In the framework of the present work, one can also expect that for very low pumping powers, where the vortices are not pinned, the scaling exponent should tend to the mean-field 2D limit of $1/2$. This limit should be tested by counting the vortex density immediately after the condensation, to avoid the effect of the decay. \tcr{The leftmost point in Fig.~\ref{fig3}(b) seems to indicate the onset of the $1/2$ scaling exponent corresponding to the low-wavevector limit of the dispersion (this slope is shown by a blue dashed line).}

\section{Conclusions}

To conclude, we have studied the polariton condensation in a honeycomb lattice. We have demonstrated that with a sufficiently fast energy relaxation, the condensation can occur at the ground state of the system (at the $\Gamma$ point of the reciprocal space). We have shown that the quantum vortices formed during the condensation via the Kibble-Zurek mechanism can be pinned to the underlying lattice, which facilitates their experimental observation in the cw regime. \tcr{We demonstrate that the presence of the lattice affects the scaling exponent of the density of the topological defects, which is found to be $0.95\pm0.05$ in numerical experiments, in agreement with the mean-field prediction of $1$ accounting for the lattice, as compared with a value of $1/2$ expected without a lattice.}

\begin{acknowledgments}
We acknowledge the support of the projects EU "QUANTOPOL" (846353), "Quantum Fluids of Light"  (ANR-16-CE30-0021), of the ANR Labex GaNEXT (ANR-11-LABX-0014), and of the ANR program "Investissements d'Avenir" through the IDEX-ISITE initiative 16-IDEX-0001 (CAP 20-25). 
\end{acknowledgments}

\bibliography{biblio}

\end{document}